\documentclass{article}
\usepackage{spconf,amsmath,graphicx}
\usepackage{cite}
\usepackage{amsthm,amsmath,amssymb,bm}
\usepackage{amsfonts}
\usepackage{booktabs}
\usepackage{url}
\usepackage{xcolor}
\usepackage{multirow}
\usepackage{makecell}
\usepackage[caption=false]{subfig}
\usepackage{float}
\usepackage[colorlinks,linkcolor=red]{hyperref}

\usepackage{bbding}
\usepackage{setspace}

\title{Synth-AC: Enhancing Audio Captioning with Synthetic Supervision} 

\name{Feiyang Xiao$^1$, Qiaoxi Zhu$^2$, Jian Guan$^{1*}$, Xubo Liu$^{3}$, Haohe Liu$^{3}$, Kejia Zhang$^1$, Wenwu Wang$^3$
\thanks{*Corresponding author. }
\thanks{This work was partly supported by the Natural Science Foundation of Heilongjiang Province under Grant No. YQ2020F010, and GHfund under Grant No. 202302026860.}
}
\address{
  $^1$College of Computer Science and Technology,  Harbin Engineering University, Harbin, 150001, China\\
  $^2$Centre for Audio, Acoustics and Vibration, University of Technology Sydney, Ultimo, NSW, Australia\\
  $^3$Centre for Vision Speech and Signal Processing, University of Surrey, Guildford, GU2 7XH, UK
  }
\begin{document}
\topmargin=0mm
%
\maketitle
\begin{abstract}
Data-driven approaches hold promise for audio captioning. However, the development of audio captioning methods can be biased due to the limited availability and quality of text-audio data. This paper proposes a SynthAC framework, which leverages recent advances in audio generative models and commonly available text corpus to create synthetic text-audio pairs, thereby enhancing text-audio representation. Specifically, the text-to-audio generation model, i.e., AudioLDM, is used to generate synthetic audio signals with captions from an image captioning dataset. Our SynthAC expands the availability of well-annotated captions from the text-vision domain to audio captioning, thus enhancing text-audio representation by learning relations within synthetic text-audio pairs. Experiments demonstrate that our SynthAC framework can benefit audio captioning models by incorporating well-annotated text corpus from the text-vision domain, offering a promising solution to the challenge caused by data scarcity. Furthermore, SynthAC can be easily adapted to various state-of-the-art methods, leading to substantial performance improvements.
\end{abstract}
\begin{keywords}
Multimodal learning, text-audio representation, audio captioning, text-to-audio generation
\end{keywords}
\section{Introduction}
\label{sec:intro}
Multimodal text-audio learning can be considered as an imitation of the hearing and natural language understanding ability of human beings \cite{baltruvsaitis2018multimodal}, which is helpful to down-stream tasks, such as audio captioning, for describing the content of the audio signal by natural language captions \cite{Drossos_2020_icassp}. In audio captioning, data-driven deep learning approaches are widely used for learning the relation between audio and text information \cite{drossos2017automated}. 
Nonetheless, the performance of these methods could be limited by the scarcity of reliable text-audio data \cite{liu2021cl4ac, liu2022leveraging}, due to the significant challenges in capturing and annotating such data.

To address the data scarcity issue of audio captioning, efforts have been made to directly increase the number of text-audio pairs by using external datasets \cite{yuan2021_t6, mei2023wavcaps, wu2023_t6a}. For example, NetEase adopted a large-scale additional training dataset and significantly improved audio captioning performance \cite{yuan2021_t6}. However, such large-scale curated datasets are generally non-public, barring their use in research and real-world applications. Recently, the large language model, e.g., ChatGPT\footnote{\url{https://openai.com/blog/chatgpt/}}, has been utilised to generate pseudo captions for audio datasets \cite{mei2023wavcaps, wu2023_t6a}. In \cite{mei2023wavcaps}, a novel audio captioning dataset, i.e., WavCaps, was created by utilising ChatGPT to convert weakly-labelled audio tags into fluent captions, which highly increases the number of text-audio data pairs and improves the performance of various down-stream tasks including audio captioning. In addition, a ChatGPT-based mixup strategy \cite{wu2023_t6a} improves the fluency and precision of the predicted captions by generating new captions from randomly paired existing captions in the Clotho dataset \cite{Drossos_2020_icassp}. 

Audio captioning is an inherently multimodal audio-language task, while recent advances often focus on enhancing audio datasets with pseudo-textual captions. Although annotated audio-text data is limited at scale, text corpus is nearly infinite to access on the web. Intuitively, an open question is raised: \textit{can we augment text corpus with pseudo-acoustic signals}? The key to this question is to take advantage of recent achievements in text-to-audio generation \cite{liu2023audioldm}. By leveraging text-to-audio generation models, we can seamlessly produce synthetic audio data for large-scale textual datasets. Text-to-audio generated synthetic data provide advantages such as having flexible control over the audio concepts and linguistic diversity. Despite its great potential, to the best of our knowledge, leveraging synthetic audio data for improving audio captioning has never been studied in the literature. 

\begin{table}[t]
    \centering
    \caption{Examples to show the latent relation between image captions and audio captions.}
    \vspace{0.5em}
    \resizebox{\columnwidth}{!}
    {
    \begin{tabular}{c|c}
        \Xhline{1pt}
        Image Caption in COCO \cite{lin2014microsoft} & Audio Caption in AudioCaps \cite{kim2019audiocaps} \\
        \Xhline{0.5pt}
        \makecell[l]{A black \textbf{car} is near someone riding\\ a bike} & \makecell[l]{A man talking and a \textbf{car} passing\\ by loudly} \\
        \Xhline{0.5pt}
        \makecell[l]{\textbf{A barking dog} looks over a ledge\\ lined with Christmas lights} & \makecell[l]{\textbf{Dog barking} and growling} \\
        \Xhline{0.5pt}
        \makecell[l]{\textbf{A cat sleeping} on a rock near a bike} & \makecell[l]{\textbf{A cat sleeps} and snores} \\
        \Xhline{1pt}
    \end{tabular}
    }
    \label{tab:examples}
    \vspace{-4mm}
\end{table}

In this paper, we propose SynthAC, a semi-supervised framework that leverages synthetic supervision from text datasets to enhance the performance of audio captioning systems. Specifically, we consider the text data from the image captioning dataset COCO \cite{lin2014microsoft}, where the well-annotated image captions describing the visual scenes have latent relations to the semantics described by audio captions used in AudioCaps dataset \cite{kim2019audiocaps}, as shown in Table~\ref{tab:examples}. In the SynthAC framework, we first generate synthetic audio data for image captions using the state-of-the-art audio generation model i.e., AudioLDM \cite{liu2023audioldm}. The synthetic audio signals with paired image captions are then used to augment the specialized audio captioning dataset (i.e., AudioCaps \cite{kim2019audiocaps}). Then, the augmented dataset is used to train the audio captioning model in a semi-supervised setting, and to enhance text-audio representation. 

We experiment on the audio captioning dataset i.e. AudioCaps \cite{kim2019audiocaps} with two state-of-the-art models, i.e., GraphAC \cite{xiao2023graph} and P-Transformer \cite{xinhao2021_t6}. Results show that SynthAC could substantially improve audio captioning performance with the variable scale of real audio-text data. Furthermore, we demonstrate that SynthAC performs on par with off-the-shelf audio captioning systems using less than half of the real data, indicating the great potential of SynthAC to mitigate the data scarcity issue in audio captioning. The generated synthetic text-audio data and caption examples are available at: {\url{https://github.com/LittleFlyingSheep/SynthAC}}.

\section{Proposed Synth-AC Method}
\label{sec:2}
\begin{figure}
    \vspace{1mm}
    \centering
    \includegraphics[width=\linewidth]{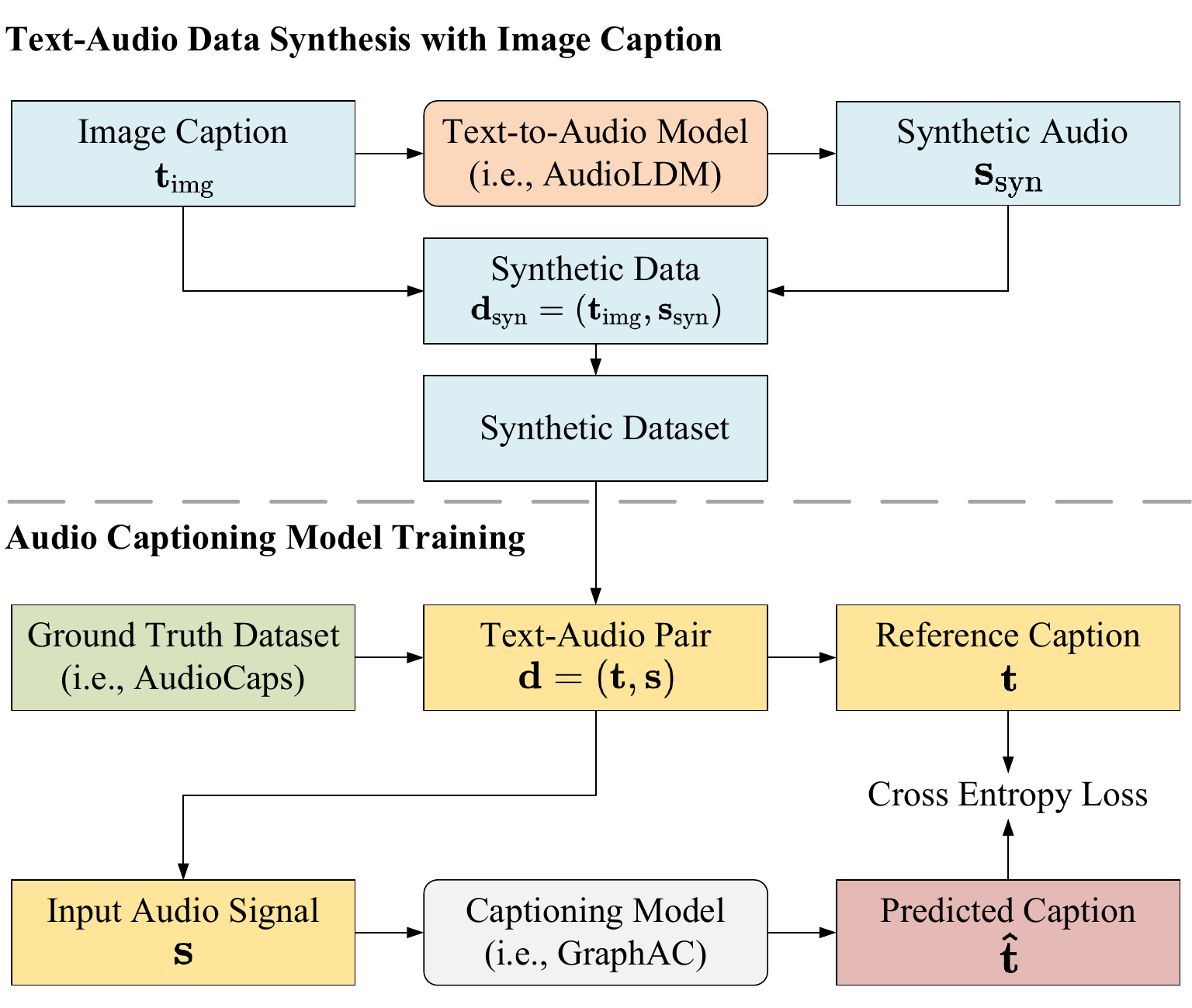}
    \vspace{-7mm}
    \caption{The proposed SynthAC framework includes two stages. (1) text-audio data synthesis with image captions, where synthetic text-audio data pairs are generated from well-annotated image captions via a text-to-audio model (i.e., AudioLDM), and (2) audio captioning model training, where the audio captioning model (i.e., GraphAC \cite{xiao2023graph}) is trained with the synthetic text-audio pairs enriched training data.
    } 
    \label{fig:workflow}
    \vspace{-4mm}
\end{figure}

This section presents the proposed audio captioning method SynthAC in detail, with the overall framework shown in Figure~\ref{fig:workflow}. The well-annotated image captions (e.g. from the COCO dataset \cite{lin2014microsoft}) are employed as the condition (input) of the latent diffusion model used in AudioLDM \cite{liu2023audioldm} for audio generation, in order to scale up the text-audio dataset AudioCaps. Then, these scaled data are employed to train the audio captioning model (i.e. GraphAC \cite{xiao2023graph}) to enhance text-audio representation learning and improve audio captioning performance. In addition to GraphAC, the proposed SynthAC framework is also adapted to other audio captioning models.

\subsection{Text-Audio Data Synthesis with Image Captions}

To obtain synthetic text-audio pairs, we use the image caption $\mathbf{t}_{\text{img}}$ as inputs to the pretrained AudioLDM model $G( \cdot) $ to generate the audio signal
\begin{equation}
    \mathbf{s}_{\text{syn}} = G(\mathbf{t}_{\text{img}}).
\end{equation}
This allows us to obtain the synthetic text-audio pair $\mathbf{d}_{\text{syn}} = (\mathbf{t}_{\text{img}}, \mathbf{s}_{\text{syn}})$. Thus, we can form a synthetic dataset $ D_{\text{syn}}$ composed of all the image captions selected from the image caption dataset, and the corresponding audio clips generated with the text-to-audio (TTA) model.

Note that the AudioLDM model we used is pretrained on AudioSet \cite{gemmeke2017audio}, AudioCaps \cite{kim2019audiocaps}, FreeSound\footnote{\url{https://freesound.org/}} and BBC Sound Effects\footnote{\url{https://sound-effects.bbcrewind.co.uk/search}} datasets, which is capable in establishing the relation between the text descriptions and acoustic scenes and events for audio generation.  Therefore, we can use such a model to generate the synthetic audio signal, with the image caption as the condition, utilising the implicit relation between audio and visual scene.

\subsection{Audio Captioning Model Training}

We employ GraphAC \cite{xiao2023graph} developed in our recent work as the audio captioning model in the proposed SynthAC framework, denoted as Synth-GraphAC. The GraphAC method employs an encoder-decoder structure, where the audio encoder is used to extract the audio feature and the text decoder is used to predict captions from the audio feature. The encoder introduces a graph attention module to capture the contextual temporal information in the audio feature extracted by a PANNs module \cite{kong2020panns}. The text decoder uses a two-layer transformer module with a Word2Vec model \cite{word2vec} for caption prediction from the audio feature.

With the synthetic data $D_{\text{syn}}$ generated with the method discussed earlier, we can augment the ground truth dataset $D_{\text{gt}}$, e.g. AudioCaps, and obtain the augmented training set 
\begin{equation}
    D_{\text{T}} = D_{\text{syn}} \cup D_{\text{gt}},
\end{equation}
where $D_{\text{T}}$ denotes the augmented training set, where each text-audio pair is denoted as $\mathbf{d} = (\mathbf{t}, \mathbf{s})$, with $\mathbf{t}$ being the caption and $\mathbf{s}$ being the corresponding audio signal.

During model training, an input audio $\mathbf{s}$ is fed into the audio captioning model to generate a predicted caption, 
\begin{equation}
    \mathbf{\hat{t}} = AC(\mathbf{s}),
\end{equation}
where $AC(\cdot)$ denotes the audio captioning model, i.e., GraphAC, and $\mathbf{\hat{t}}$ denotes the predicted caption. 
Then the cross entropy (CE) loss function with label smoothing \cite{xiao2023graph, xinhao2021_t6} is used to optimize the audio captioning model
\begin{equation}
    \mathcal{L} = \text{CE}(\mathbf{\hat{t}}, \mathbf{t}).
\end{equation}
With the synthetic data to enhance the text-audio representation, our proposed SynthAC can further improve the performance of the audio captioning models.

In addition to GraphAC, the proposed SynthAC method can be easily adapted to different audio captioning models to improve their performance. In the following experiment, another audio captioning model, i.e., P-Transformer \cite{xinhao2021_t6} is also employed in the proposed SynthAC framework, denoted as Synth-P-Transformer, to demonstrate the effectiveness of our proposed method further, as detailed in Section~\ref{subsec:comparison}.

\section{Experiments}
\label{sec:experiment}
\subsection{Experimental Setup}

\textbf{Dataset: }
For text-audio data synthesis, we employ the well-annotated image captions from the widely used text-visual application dataset, i.e., COCO \cite{lin2014microsoft}, which provides 414,113 high quality manually annotated image captions to describe visual scenes. Here, 25,000 individual image captions are randomly selected from the COCO dataset, which are used as prompts of the AudioLDM model to obtain a total number of 25,000 synthetic text-audio pairs to enhance the text-audio representation for model training.

For audio captioning model training, we employ the widely used audio captioning dataset, i.e., AudioCaps \cite{kim2019audiocaps}, as the ground truth dataset. The development and validation splits of AudioCaps with 51,744 text-audio pairs are combined for model training, following \cite{xinhao2021_t6, xiao2022local, xiao2023graph}, and the evaluation split is used for evaluation. The sampling rate of audio signals is 16kHz, as the default setting of AudioLDM.

\vspace{1.5mm}
\noindent \textbf{Implementation Details: } In data synthesis stage, we employ the ``audioldm-l-full" version of AudioLDM model\footnote{\url{https://github.com/haoheliu/AudioLDM}} for synthetic audio generation. The length of synthetic audio is 10 seconds, consistent with those in AudioCaps. During the model training, the batch size is set as 16 for both Synth-GraphAC and Synth-P-Transformer. The AdamW optimiser \cite{loshchilov2017fixing} with a learning rate of 0.001 is used for model training. SpecAugment is used to enhance the generalisation for audio captioning models following \cite{xiao2022local, xiao2023graph, xinhao2021_t6}.

\vspace{1.5mm}
\noindent \textbf{Evaluation Metrics: }For performance evaluation, BLEU$_n$, ROUGE$_l$, METEOR, CIDE$_r$, SPICE, SPIDE$_r$ and SPIDE$_r$-FL, are employed as evaluation metrics in our experiments, following \cite{guan2023_t6, Drossos_2020_icassp}. BLEU$_n$, ROUGE$_l$ and METEOR measure the matching degree between the prediction and ground truth caption on word level \cite{xinhao2021_t6}. CIDE$_r$ measures the fluency of the caption \cite{vedantam2015cider}. SPICE measures the semantic proposition between the predicted caption and the reference caption \cite{SPICE}. SPIDE$_r$ is the average value of the CIDE$_r$ and SPICE metrics, which balances the measure of the fluency and the semantic information in the caption evaluation \cite{liu2017improved}. SPIDE$_r$-FL is a recently proposed metric introducing fluency error based penalty for improving the robustness of the evaluation \cite{FENSE2022}.

\subsection{Effect of the Proposed SynthAC}
\label{subsec:comparison}

To show the effectiveness of the proposed SynthAC, we conduct experiments to compare the proposed method with audio captioning methods only trained with the ground truth dataset, including GPT-Similar \cite{koizumi2020audio}, TopDown-AlignedAtt \cite{kim2019audiocaps}, P-Transformer \cite{xinhao2021_t6}, GraphAC \cite{xiao2023graph} and P-LocalAFT \cite{xiao2022local}. GPT-Similar is a typical captioning method using similar caption retrieval to predict caption. TopDown-AlignedAtt is the study that proposed AudioCaps \cite{kim2019audiocaps}. P-Transformer, GraphAC and P-LocalAFT are state-of-the-art methods in audio captioning. To validate the proposed method, P-Transformer and GraphAC are employed as the captioning model in the SynthAC framework, denoted as Synth-P-Transformer and Synth-GraphAC, respectively. 

Table~\ref{tab:performance} shows that, with the enhanced text-audio relations by the synthetic data, the proposed SynthAC framework can improve the audio captioning models' performance, which can be observed from the comparison between Synth-GraphAC and GraphAC, as well as the comparison between Synth-P-Transformer and P-Transformer. In addition, the examples of the predicted captions using P-Transformer and Synth-P-Transformer are provided in Table~\ref{tab:illustration} to show the improvement on captioning results with our SynthAC. 

\begin{table*}[t]
    \centering
    \caption{Performance comparison on the evaluation split of AudioCaps.}
    \vspace{0.5em}
    \resizebox{\textwidth}{!}
    {
    \begin{tabular}{ccccccccccc}
        \toprule
        Model & BLEU$_1$ & BLEU$_2$ & BLEU$_3$ & BLEU$_4$ & ROUGE$_l$ & METEOR & CIDE$_r$ & SPICE & SPIDE$_r$ & SPIDE$_r$-FL \\
        \midrule
        GPT-Similar \cite{koizumi2020audio} & 63.8 & 45.8 & 31.8 & 20.4 & 43.4 & 19.9 & 50.3 &  13.9 & 32.1 & - \\
        TopDown-AlignedAtt \cite{kim2019audiocaps} & 61.4 & 44.6 & 31.7 & 21.9 & 45.0 & 20.3 & 59.3 & 14.4 & 36.9 & - \\
        P-LocalAFT \cite{xiao2022local} & 66.0 & 47.9 & 34.6 & 24.6 & 46.4 & 22.3 & 64.1 & 16.6 & 40.4 & 40.0 \\
        \midrule
        P-Transformer \cite{xinhao2021_t6} & 53.4 & 38.9 & 27.1 & 18.0 & 44.2 & 21.5 & 57.7 & 16.6 & 37.1 & 35.9 \\
        \textbf{Synth-P-Transformer} & \underline{\textbf{67.7}} & \underline{\textbf{49.9}} & \underline{\textbf{36.0}} & \underline{\textbf{25.1}} & \underline{\textbf{46.8}} & \underline{\textbf{22.7}} & \textbf{63.9} & \underline{\textbf{16.7}} & \textbf{40.3} & \textbf{39.4} \\
        \midrule
        GraphAC \cite{xiao2023graph} & 64.5 & 47.8 & 34.3 & 23.7 & 46.1 & \textbf{22.4} & 64.4 & \underline{\textbf{16.7}} & 40.5 & 39.3 \\
        \textbf{Synth-GraphAC} & \textbf{66.5} & \textbf{48.7} & \textbf{35.2} & \textbf{24.7} & \textbf{46.4} & \textbf{22.4} & \underline{\textbf{65.6}} & 16.5 & \underline{\textbf{41.0}} & \underline{\textbf{40.4}} \\
        \bottomrule
    \end{tabular}
    }
    \label{tab:performance}
    \vspace{-3mm}
\end{table*}

\begin{table}[t]
    \vspace{-2mm}
    \centering
    \caption{Illustration for audio captioning with or without SynthAC framework.}
    \vspace{0.5em}
    \resizebox{\linewidth}{!}
    {
    \begin{tabular}{c|l|l}
        \Xhline{1pt}
                           & \makecell[c]{Model}                & \makecell[c]{Audio Caption} \\
        \Xhline{0.5pt}
        \multirow{4}{*}{1} & Reference caption                  & A helicopter \textbf{blades running} \\
                           \Xcline{2-3}{0.5pt}
                           & P-Transformer \cite{xinhao2021_t6} & \makecell[l]{A helicopter machine \textit{gun fires}\\ \textit{rapidly}} \\
                           \Xcline{2-3}{0.5pt}
                           & \textbf{Synth-P-Transformer}       & Helicopter \textbf{blades spinning} \\
        \Xhline{0.5pt}
        \multirow{4}{*}{2} & Reference                          & \makecell[l]{A man talking \textbf{followed by} a toilet\\ flushing} \\
                           \Xcline{2-3}{0.5pt}
                           & P-Transformer \cite{xinhao2021_t6} & \makecell[l]{A man speaking a toilet flushing} \\
                           \Xcline{2-3}{0.5pt}
                           & \textbf{Synth-P-Transformer}       & \makecell[l]{A man speaks \textbf{followed by} a toilet\\ flushing} \\
        \Xhline{0.5pt}
        \multirow{4}{*}{3} & Reference caption                  & \makecell[l]{A woman speaks with some rattling\\ and some \textbf{spraying}} \\
                           \Xcline{2-3}{0.5pt}
                           & P-Transformer \cite{xinhao2021_t6} & An adult female is speaking \\
                           \Xcline{2-3}{0.5pt}
                           & \textbf{Synth-P-Transformer}       & \makecell[l]{A woman speaking followed by\\ \textbf{spraying}} \\
        \Xhline{1pt}
    \end{tabular}
    }
    \label{tab:illustration}
    \vspace{-4mm}
\end{table}

From Table~\ref{tab:illustration}, we can see that the ``blades running" is wrongly interpreted as ``gun fires rapidly" by P-Transformer, whereas Synth-P-Transformer precisely predicts this concept with the enhanced text-audio representation, as illustrated in example 1. For example 2, Synth-P-Transformer precisely describes the contextual information (i.e., ``followed by"), and obtains exactly the same predicted caption as the reference. Regarding example 3, the acoustic event ``spraying" is missed in the prediction of P-Transformer but predicted correctly by Synth-P-Transformer. The improved predictions can be also seen in terms of the word precision metric BLEU$_n$, the fluency metric CIDE$_r$, and the kernel semantic metric SPIDE$_r$-FL, as shown in Table~\ref{tab:performance}.

Meanwhile, the proposed SynthAC based methods also outperform other methods, demonstrating the effectiveness of the proposed SynthAC framework. Furthermore, the comparison results show that, with our proposed SynthAC framework, we can use the well-annotated captions from the text-vision multimodal domain to enhance the text-audio representation learning in multimodal audio captioning and reduce the cost of obtaining text-audio data.

\subsection{Performance Evaluation with Different Amounts of Real Data}
We evaluate the proposed SynthAC (i.e., Synth-P-Transformer) by using different amounts of ground truth data for model training, i.e. 12.5\%, 25\%, 37.5\% and 50\% of AudioCaps dataset, respectively. We also compare the performance of the model, trained with and without synthetic data. The results are shown in Table \ref{tab:ablation}. 

{Table~\ref{tab:ablation} shows that the proposed SythnAC can significantly improve the captioning performance across the different amounts of ground truth data used for model training, especially with a very limited amount of data, i.e., 12.5\% of ground truth. With the enhanced text-audio relation by adopting our proposed SynthAC framework, Synth-P-Transformer outperforms the GPT-Similar with the complete training data in Table \ref{tab:performance}. In addition, the Synth-P-Transformer using only 37.5\% ground truth data has better captioning performance than the P-Transformer with the complete training data in Table \ref{tab:performance}. The results further demonstrate the effectiveness of our proposed SynthAC framework, which provides a  solution for audio captioning with only limited text-audio data pairs. Moreover, the above results verify that properly using well-annotated textual information from text corpus in a multimodal learning task (i.e., text-vision domain) can benefit another multimodal learning task (i.e., audio captioning) suffering from data scarcity.}

\begin{table}[t]
    \vspace{-2mm}
    \centering
    \caption{Performance of Synth-P-Transformer trained with different amounts of the ground truth dataset (AudioCaps) and with or without synthetic data.}
    \vspace{0.5em}
    \resizebox{\columnwidth}{!}
    {
    \begin{tabular}{cccccc}
        \toprule
        \makecell[c]{Real Data\\Percentage} & \makecell[c]{Synthetic\\ Data} & CIDE$_r$ & SPICE & SPIDE$_r$ & SPIDE$_r$-FL \\
        \midrule
        \multirow{2}{*}{\makecell[c]{12.5\% }}   & \XSolid    & 49.9          & 14.3          & 32.1          & 29.9 \\
                                             & \Checkmark & \textbf{55.2} & \textbf{15.1} & \textbf{35.2} & \textbf{33.1} \\
        \midrule
        \multirow{2}{*}{\makecell[c]{25.0\% }}   & \XSolid    & 56.1          & 13.6          & 34.8          & 33.8 \\
                                             & \Checkmark & \textbf{58.4} & \textbf{15.5} & \textbf{36.9} & \textbf{35.0} \\
        \midrule
        \multirow{2}{*}{\makecell[c]{37.5\%}}   & \XSolid    & 58.1          & 15.1          & 36.6          & 35.3 \\
                                             & \Checkmark & \textbf{61.1} & \textbf{15.9} & \textbf{38.5} & \textbf{37.6} \\
        \midrule
        \multirow{2}{*}{\makecell[c]{50.0\% }}   & \XSolid    & 57.6          & 16.2          & 36.9          & 34.4 \\
                                             & \Checkmark & \textbf{63.8} & \textbf{16.7} & \textbf{40.2} & \textbf{38.3} \\
        \bottomrule
    \end{tabular}
    }
    \label{tab:ablation}
    \vspace{-4mm}
\end{table}

\section{Conclusion}

We have presented an audio captioning framework with synthetic supervision, leveraging well-annotated text-vision captions in the image captioning dataset and text-to-audio generation model to enhance the learning of text-audio representation and improve audio captioning performance.  Experiments show the proposed method's effectiveness, which offers improved performance compared to the baseline methods, can be easily adapted to various state-of-the-art methods with substantial performance improvements, and can maintain performance with a much-reduced amount of actual text-audio data, offering a promising solution to the challenge of data scarcity.

\begin{spacing}{0.89}
    \bibliographystyle{IEEEtran}
    \bibliography{strings}

\begin{thebibliography}{10}
\providecommand{\url}[1]{#1}
\csname url@samestyle\endcsname
\providecommand{\newblock}{\relax}
\providecommand{\bibinfo}[2]{#2}
\providecommand{\BIBentrySTDinterwordspacing}{\spaceskip=0pt\relax}
\providecommand{\BIBentryALTinterwordstretchfactor}{4}
\providecommand{\BIBentryALTinterwordspacing}{\spaceskip=\fontdimen2\font plus
\BIBentryALTinterwordstretchfactor\fontdimen3\font minus
  \fontdimen4\font\relax}
\providecommand{\BIBforeignlanguage}[2]{{%
\expandafter\ifx\csname l@#1\endcsname\relax
\typeout{** WARNING: IEEEtran.bst: No hyphenation pattern has been}%
\typeout{** loaded for the language `#1'. Using the pattern for}%
\typeout{** the default language instead.}%
\else
\language=\csname l@#1\endcsname
\fi
#2}}
\providecommand{\BIBdecl}{\relax}
\BIBdecl

\bibitem{baltruvsaitis2018multimodal}
T.~Baltru{\v{s}}aitis, C.~Ahuja, and L.-P. Morency, ``Multimodal machine
  learning: A survey and taxonomy,'' \emph{IEEE Trans. Pattern Anal. Mach.
  Intell.}, vol.~41, no.~2, pp. 423--443, 2018.

\bibitem{Drossos_2020_icassp}
K.~Drossos, S.~Lipping, and T.~Virtanen, ``Clotho: An audio captioning
  dataset,'' in \emph{Proc. IEEE Int. Conf. Acoust., Speech, Signal Process.},
  2020, pp. 736--740.

\bibitem{drossos2017automated}
K.~Drossos, S.~Adavanne, and T.~Virtanen, ``Automated audio captioning with
  recurrent neural networks,'' in \emph{Proc. IEEE Workshop Appl. Signal
  Process. Audio Acoust.}, 2017, pp. 374--378.

\bibitem{liu2021cl4ac}
X.~Liu, Q.~Huang, X.~Mei, T.~Ko, H.~Tang, M.~D. Plumbley, and W.~Wang,
  ``{CL4AC}: A contrastive loss for audio captioning,'' in \emph{Proc.
  Detection Classification Acoust. Scenes Events Workshop}, 2021, pp. 196--200.

\bibitem{liu2022leveraging}
X.~Liu, X.~Mei, Q.~Huang, J.~Sun, J.~Zhao, H.~Liu, M.~D. Plumbley, V.~Kilic,
  and W.~Wang, ``Leveraging pre-trained {BERT} for audio captioning,'' in
  \emph{Proc. European Signal Proces. Conf.}, 2022, pp. 1145--1149.

\bibitem{yuan2021_t6}
W.~Yuan, Q.~Han, D.~Liu, X.~Li, and Z.~Yang, ``The {DCASE} 2021 challenge task
  6 system: Automated audio captioning with weakly supervised pre-traing and
  word selection methods,'' Detection Classification Acoust. Scenes Events
  Challenge, Tech. Rep., 2021.

\bibitem{mei2023wavcaps}
X.~Mei, C.~Meng, H.~Liu, Q.~Kong, T.~Ko, C.~Zhao, M.~D. Plumbley, Y.~Zou, and
  W.~Wang, ``{WavCaps}: A {ChatGPT}-assisted weakly-labelled audio captioning
  dataset for audio-language multimodal research,'' \emph{arXiv preprint
  arXiv:2303.17395}, 2023.

\bibitem{wu2023_t6a}
S.-L. Wu, X.~Chang, G.~Wichern, J.-w. Jung, F.~Germain, J.~L. Roux, and
  S.~Watanabe, ``{BEATs}-based audio captioning model with {INSTRUCTOR}
  embedding supervision and {ChatGPT} mix-up,'' Detection Classification
  Acoust. Scenes Events Challenge, Tech. Rep., 2023.

\bibitem{liu2023audioldm}
H.~Liu, Z.~Chen, Y.~Yuan, X.~Mei, X.~Liu, D.~Mandic, W.~Wang, and M.~D.
  Plumbley, ``{AudioLDM}: Text-to-audio generation with latent diffusion
  models,'' in \emph{Proc. Int. Conf. Machin. Learn.}, 2023.

\bibitem{lin2014microsoft}
T.-Y. Lin, M.~Maire, S.~Belongie, J.~Hays, P.~Perona, D.~Ramanan,
  P.~Doll{\'a}r, and C.~L. Zitnick, ``Microsoft {COCO}: Common objects in
  context,'' in \emph{Proc. European Conf. Comput. Vision}, 2014, pp. 740--755.

\bibitem{kim2019audiocaps}
C.~D. Kim, B.~Kim, H.~Lee, and G.~Kim, ``{AudioCaps}: Generating captions for
  audios in the wild,'' in \emph{Proc. N. Am. Chapter Assoc. Comput.
  Linguistics: Hum. Lang. Technol.}, 2019, pp. 119--132.

\bibitem{xiao2023graph}
F.~Xiao, J.~Guan, Q.~Zhu, and W.~Wang, ``Graph attention for automated audio
  captioning,'' \emph{IEEE Signal Process. Lett.}, vol.~30, pp. 413--417, 2023.

\bibitem{xinhao2021_t6}
X.~Mei, Q.~Huang, X.~Liu, G.~Chen, J.~Wu, Y.~Wu, J.~Zhao, S.~Li, T.~Ko, H.~L.
  Tang, X.~Shao, M.~D. Plumbley, and W.~Wang, ``An encoder-decoder based audio
  captioning system with transfer and reinforcement learning,'' in \emph{Proc.
  Detection Classification Acoust. Scenes Events Workshop}, 2021.

\bibitem{gemmeke2017audio}
J.~F. Gemmeke, D.~P. Ellis, D.~Freedman, A.~Jansen, W.~Lawrence, R.~C. Moore,
  M.~Plakal, and M.~Ritter, ``{AudioSet}: An ontology and human-labeled dataset
  for audio events,'' in \emph{Proc. IEEE Int. Conf. Acoust., Speech, Signal
  Process.}, 2017, pp. 776--780.

\bibitem{kong2020panns}
Q.~Kong, Y.~Cao, T.~Iqbal, Y.~Wang, W.~Wang, and M.~D. Plumbley, ``{PANNs}:
  Large-scale pretrained audio neural networks for audio pattern recognition,''
  \emph{IEEE-ACM Trans. Audio Speech Lang.}, vol.~28, pp. 2880--2894, 2020.

\bibitem{word2vec}
G.~C. Tomas~Mikolov, Kai~Chen, ``Efficient estimation of word representations
  in vector space,'' in \emph{Proc. Int. Conf. Learn. Represent.}, 2013.

\bibitem{xiao2022local}
F.~Xiao, J.~Guan, H.~Lan, Q.~Zhu, and W.~Wang, ``Local information assisted
  attention-free decoder for audio captioning,'' \emph{IEEE Signal Process.
  Lett.}, vol.~29, pp. 1604--1608, 2022.

\bibitem{loshchilov2017fixing}
I.~Loshchilov and F.~Hutter, ``Fixing weight decay regularization in adam.
  arxiv 2017,'' \emph{arXiv preprint arXiv:1711.05101}, vol.~7, 2017.

\bibitem{guan2023_t6}
F.~Xiao, Q.~Zhu, H.~Lan, W.~Wang, and J.~Guan, ``Ensemble systems with
  contrastive language-audio pretraining and attention-based audio features for
  audio captioning and retrieval,'' Detection Classification Acoust. Scenes
  Events Challenge, Tech. Rep., 2023.

\bibitem{vedantam2015cider}
R.~Vedantam, C.~Lawrence~Zitnick, and D.~Parikh, ``{CIDEr}: Consensus-based
  image description evaluation,'' in \emph{Proc. IEEE Comput. Soc. Conf.
  Comput. Vision Pattern Recognit.}, 2015, pp. 4566--4575.

\bibitem{SPICE}
P.~Anderson, B.~Fernando, M.~Johnson, and S.~Gould, ``{SPICE}: Semantic
  propositional image caption evaluation,'' in \emph{Proc. European Conf.
  Comput. Vision}, 2016, pp. 382--398.

\bibitem{liu2017improved}
S.~Liu, Z.~Zhu, N.~Ye, S.~Guadarrama, and K.~Murphy, ``Improved image
  captioning via policy gradient optimization of {SPIDEr},'' in \emph{Proc.
  IEEE Int. Conf. Comput. Vision}, 2017, pp. 873--881.

\bibitem{FENSE2022}
Z.~Zhou, Z.~Zhang, X.~Xu, Z.~Xie, M.~Wu, and K.~Q. Zhu, ``Can audio captions be
  evaluated with image caption metrics?'' in \emph{Proc. IEEE Int. Conf.
  Acoust., Speech, Signal Process.}, 2022, pp. 981--985.

\bibitem{koizumi2020audio}
Y.~Koizumi, Y.~Ohishi, D.~Niizumi, D.~Takeuchi, and M.~Yasuda, ``Audio
  captioning using pre-trained large-scale language model guided by audio-based
  similar caption retrieval,'' \emph{arXiv preprint arXiv:2012.07331}, 2020.

\end{thebibliography}
\end{spacing}

\end{document}